\documentclass[journal]{IEEEtran}
\usepackage{comment}
\usepackage{csquotes}
\usepackage{multirow}
\usepackage{multicol}
\usepackage{amssymb}
\usepackage{amsthm}
\usepackage{amsmath}

\usepackage{enumitem}
\usepackage{mleftright}

\usepackage{array}
\newcolumntype{P}[1]{>{\centering\arraybackslash}p{#1}}

\usepackage{xcolor}
\usepackage{bm}
\usepackage{graphicx}

\usepackage[hidelinks]{hyperref}
\hypersetup{
    colorlinks=false,
    linkcolor=blue,
    filecolor=magenta,
    urlcolor=cyan,
}

\usepackage{cite}

\usepackage{pdflscape}
\usepackage{booktabs}
\usepackage{caption}

\usepackage{fancyhdr}
\usepackage[landscape]{}
\usepackage{ragged2e}
\usepackage{ltablex, makecell}
\usepackage{tabularx}

\newcommand{\tabitem}{\textbullet~~}

\begin{document}

%
\title{Interaction Graphs for Cascading Failure Analysis in Power Grids: A Survey}
%
%
%


\author{Upama Nakarmi,~\IEEEmembership{Student Member,~IEEE}, Mahshid Rahnamay-Naeini,~\IEEEmembership{Member,~IEEE},
        Md Jakir Hossain,~\IEEEmembership{Student Member,~IEEE}, Md Abul Hasnat,~\IEEEmembership{Student Member,~IEEE}
\thanks{The authors are with the Department
of Electrical Engineering, University of South Florida, Tampa,
FL, 33620.\protect\\
E-mail addresses: unakarmi@mail.usf.edu, mahshidr@usf.edu, mdjakir@mail.usf.edu, hasnat@mail.usf.edu
}
\thanks{ }}


\maketitle

\begin{abstract}
Understanding and analyzing cascading failures in power grids have been the focus of many researchers for years. 
However, the complex interactions among the large number of components in these systems and their contributions to cascading failures are not yet completely understood.
Therefore, various techniques have been developed and used to model and analyze the underlying interactions among the components of the power grid with respect to cascading failures. 
Such methods are important to reveal the essential information that may not be readily available from power system physical models and topologies.
In general, the influences and interactions among the components of the system may occur both locally and at distance due to the physics of electricity governing the power flow dynamics as well as other functional and cyber dependencies among the components of the system. 
To infer and capture such interactions, data-driven approaches or techniques based on the physics of electricity have been used to develop graph-based models of interactions among the components of the power grid. 
In this survey, various methods of developing interaction graphs as well as studies on the reliability and cascading failure analysis of power grids using these graphs have been reviewed.
\end{abstract} 



%
\IEEEpeerreviewmaketitle

\section{Introduction}
\label{Introduce}

\subsection{An Overview of the Review and Overall Significance}

Cascading failures in power grids are successive interdependent failures of components in the system, which are usually initiated by few outages due to internal or exogenous disturbances and are propagated in a relatively short period of time leading to large blackouts \cite{baldick2008initial, hines2009cascading}. Examples of blackouts that resulted from cascading failures are the case of US Northeast blackout in 2003 \cite{force2004us}, Italian blackout in 2003 \cite{berizzi2004italian}, Brazilian blackout in 2009 \cite{ilho2010brazilian}, and Indian blackout in 2012 \cite{lai2013investigation}. 
While large blackouts are infrequent, their occurrence still has substantial risks associated with the significant economic losses and social impacts that they cause. 
Understanding and mitigating cascading failures in power systems remain a challenge due to the large size of these systems as well as complex and sometimes hidden interactions among the components.
Various studies and models have been developed to understand and control these complex phenomena including methods based on power system simulation \cite{dobson2007complex, hines2011topological}, deterministic analytical models \cite{demarco2001phase}, probabilistic models \cite{rahnamay2014stochastic, ren2008using, dobson2012estimating}, and graph-based models \cite{ hines2013dual, wei2018novel, wei2018complex, wei2019identification, wang2019modeling, zang2019adjacent, yang2019graph, wei2019electrical, qi2015interaction, qi2018efficient, ju2015simulation, ju2017multi, chen2019mitigation, luo2017identify, hines2017cascading, zhou2018can, nakarmi2018analyzing, nakarmi2019critical, zhou2019markovian, wu2007state, chang2011performance, li2018temporal, li2019quantify, li2020state, liu2020identifying,  zhu2013risk, sun2014revealing, zhu2014resilience, kaplunovich2014statistical, yang2017vulnerability, guo2017monotonicity, guo2018failure, ma2018fast, ma2017speeding, ma2017fast, zio2012analyzing, poudel2018electrical, hines2008centrality, blumsack2009defining, cuffe2015visualizing, bompard2009analysis, wang2011electrical, bompard2012extended, arianos2009power, koc2014structural, kocc2014impact, wang2015network, caro2015voltage, caro2015centrality, cotilla2012comparing, cotilla2013multi}. 
For a survey of various methods for studying cascading failures, see \cite{vaiman2012risk, cuadra2015critical, guo2017critical, abedi2018review, haes2019survey, nakarmi2019interaction}.
Each of these approaches shed light on different aspects of these phenomena.

Among these categories of approaches, graph-based methods have attracted a lot of attention due to the simplicity of the models and their ability to describe the propagation behavior of the failures on the graph of the system \cite{buldyrev2010catastrophic, crucitti2004model}. Many initial graph-based models were developed based on the physical topology of the power system, where the connections among the nodes represent the actual physical connections among the components of the system \cite{albert2004structural, holmgren2006using}. However, the studies in \cite{hines2010topological, cotilla2012comparing} showed the lack of strong connection between the physical topology of the system and failure propagation in cascading failures in power grids.
In general, influences and interactions among the components of the system during cascade process may occur both locally and at distance due to the physics of electricity governing the power flow dynamics as well as other functional and cyber dependencies among the components of the system. For instance, historical as well as simulation data verify that failure of a critical transmission line in the power grid may cause overload/failure of another transmission line that may or may not be topologically close.
Therefore, graph models based on the physical topology of the system are not adequate in describing the propagation behavior of failures in power grids. 
Hence, new methods are emerging to reveal the complex and hidden interactions that may not be readily available from physical topology of the power system. 
These new approaches are focused on extracting and modeling the underlying graph of interactions among the components of the system. 
While the focus of this survey is on graph-based methods, other modeling approaches such as probabilistic, risk analysis-based, and agent-based approaches \cite{Kr_ger_2011} can also be used for modeling interactions among the components of the system.

\subsection{Review Methodology}
\label{ReviewMethodology}
In this survey, various techniques for building the {\it interaction graphs} are reviewed. While the main focus of this survey is on the methods for constructing interaction graphs, reliability studies, and analyses performed using such graphs are also briefly discussed.
The benefit of interaction graphs is that the interactions among the components are topologically local, which simplifies the study and analysis of propagation behavior of failures and properties/roles of various components in the system during the cascade process.
Note that as cascading failures are attributed to the transmission network of power grids, the focus of these studies are mainly on the transmission network. 
Moreover, the majority of  studies based on the interaction graphs are focused on cascading failure analyses in power grids; however, interaction graphs can be used for other applications in power grids (e.g., analyzing reliability to targeted attacks) or even other networked systems such as transportation networks \cite{nakarmi2017towards, nakarmi2017influence}.

The earliest research study in finding interactions between components can be traced back to the 1989 study in \cite{lagonotte1989structural}, in which interactions between the buses in the system represented a measure of electrical distance between components based on changes in voltage magnitude sensitivities.
While~the power grid analyzed in this study was not modeled as a graph, the concept in this study has been widely adapted by numerous graph-based models to represent interactions between components.
In~this survey, methods for constructing interaction graphs are broadly categorized into two main classes: {\it data-driven} approaches and {\it electric distance-based} approaches. As the name implies, the data-driven approaches for building interaction graphs rely on data collected from the system (historical and real data or simulation data) for inferring and characterizing interactions among the components of the system.
Further, three categories are defined for data-driven interaction graphs based on the method used for analyzing the data.
These include: (1) methods based on outage sequence analysis \cite{hines2013dual, wei2018novel, wei2018complex, wei2019identification, wang2019modeling, zang2019adjacent, yang2019graph, wei2019electrical, qi2015interaction, qi2018efficient, ju2015simulation, ju2017multi, luo2017identify, chen2019mitigation, hines2017cascading, zhou2018can, nakarmi2018analyzing, nakarmi2019critical, zhou2019markovian, wu2007state, chang2011performance, li2018temporal, li2019quantify, li2020state, liu2020identifying}, (2)~risk-graph methods \cite{zhu2013risk, sun2014revealing, zhu2014resilience, kaplunovich2014statistical}, and (3) correlation-based methods \cite{nakarmi2018analyzing, nakarmi2019critical, yang2017vulnerability, hasnat2019}.
The category of outage sequence analysis is further divided into four sub-categories including (i) consecutive failure-based methods \cite{hines2013dual, wei2018novel, wei2018complex, wei2019identification, wang2019modeling, zang2019adjacent, yang2019graph, wei2019electrical}, (ii) generation-based methods \cite{qi2015interaction, qi2018efficient, ju2015simulation, ju2017multi, luo2017identify, chen2019mitigation}, (iii) influence-based methods \cite{hines2017cascading, zhou2018can, nakarmi2018analyzing, nakarmi2019critical}, and (iv) multiple and simultaneous failure-based methods \cite{zhou2019markovian, wu2007state, chang2011performance, li2018temporal, li2019quantify, li2020state, liu2020identifying}. This novel taxonomy is used to classify thirty detailed research studies, including conference and journal publications, in the data-driven category into various subcategories.

\begin{figure*}[t]
\centering
    \includegraphics[width=6in,height=2.5in]{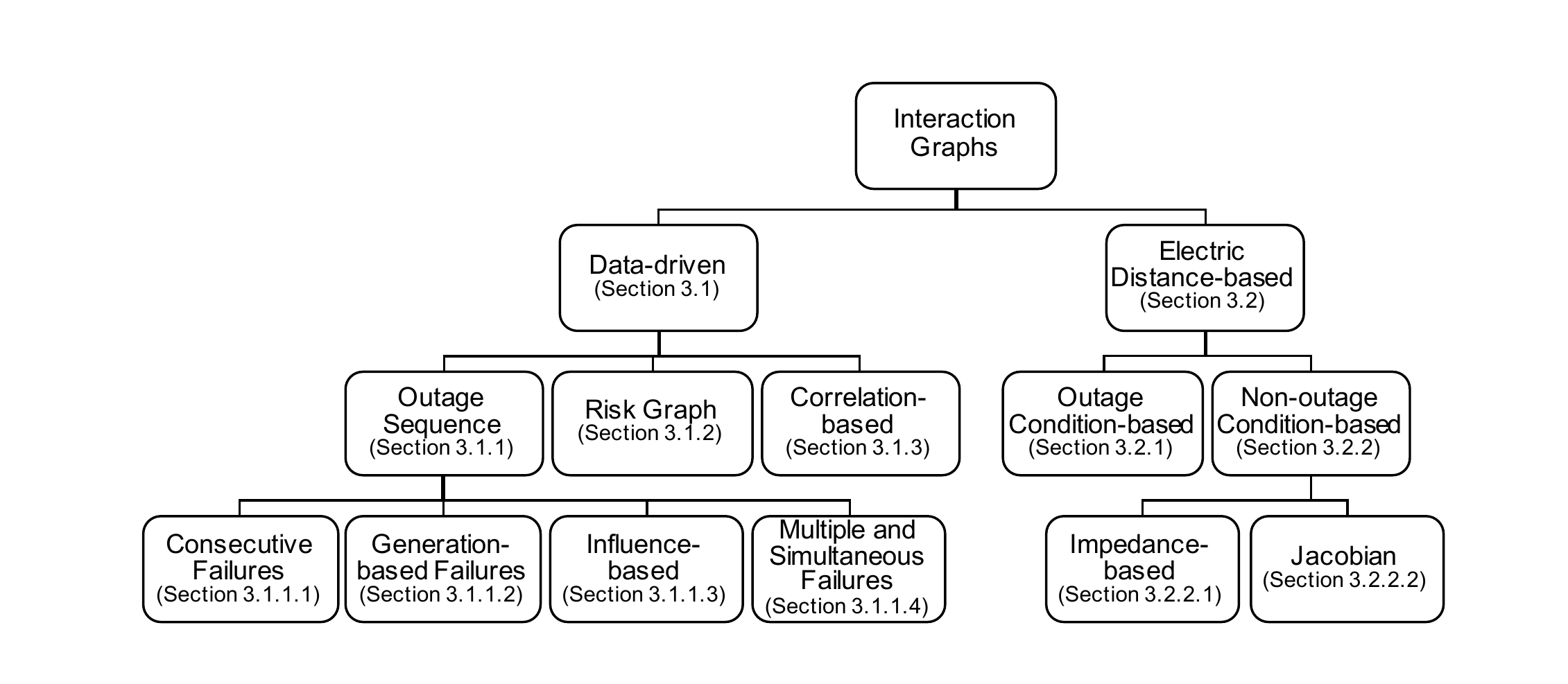}
\caption{Taxonomy of methods for constructing interaction graphs with section numbers shown for the categories and sub-categories.}
\label{Fig1:Taxonomy}
\end{figure*}

On the other hand, electric distance-based approaches exploit properties based on the physics of power and electricity governed by Kirchoff's laws to define interactions among the components. 
Thus, the interactions are represented by {\it electrical distances}, which illustrates the properties of the electrical interactions based on power flows among the components.
Two sub-categories are defined for electric distance-based interaction graphs based on the power grid conditions that are considered for creating the graphs.
These include: (1) methods that define the interactions based on changes in the power flow due to changes in physical attributes of components caused by outage conditions \cite{guo2018failure, guo2017monotonicity, ma2018fast, ma2017speeding, ma2017fast} and (2) methods that define the interactions among components during normal or non-outage operating conditions \cite{zio2012analyzing, poudel2018electrical, hines2008centrality, blumsack2009defining, cuffe2015visualizing, bompard2009analysis, wang2011electrical, koc2014structural, kocc2014impact, wang2015network, bompard2012extended, arianos2009power, caro2015voltage, caro2015centrality, cotilla2012comparing, cotilla2013multi, lagonotte1989structural}. 
The category of defining interactions during non-outage operating conditions is classified into two sub-categories: (i) impedance-based methods, which define interactions by considering a single impedance measure among the components connected over multiple paths  \cite{zio2012analyzing, poudel2018electrical, hines2008centrality, blumsack2009defining, cuffe2015visualizing, bompard2009analysis, wang2011electrical, koc2014structural, kocc2014impact, wang2015network, arianos2009power, bompard2012extended}
and (ii) sensitivities in components' states due to changes in voltage magnitudes and voltage phase angles \cite{caro2015voltage, caro2015centrality, cotilla2012comparing, cotilla2013multi, lagonotte1989structural}. 
This novel taxonomy is used to classify twenty-one detailed research studies, including conference and journal publications, in the electric distance-based category into various subcategories.
Figure \ref{Fig1:Taxonomy} shows the taxonomy of the reviewed methods for constructing various types of interaction graphs. 
The figure also specifies section numbers for the categories and subcategories, in which the methods have been discussed.
Research studies included in this review have been found using databases such as IEEE Explore, Elsevier, AIP (American Institute of Physics) Publishing, Springer, APS (American Physical Society) Physics, and MDPI. Most studies in this review are fairly recent and have been published in the past decade.

In addition to the review of various methods for constructing interaction graphs, various reliability analysis and studies performed using these graphs are also briefly reviewed.
Some studies of interaction graphs are focused on identifying critical components in the cascade process of power grids \cite{wei2018novel, wei2018complex, wei2019identification, wang2019modeling, zang2019adjacent, yang2019graph, wei2019electrical, qi2015interaction, qi2018efficient, ju2015simulation, ju2017multi, hines2017cascading, zhou2018can, nakarmi2018analyzing, nakarmi2019critical, zhou2019markovian, zhu2013risk, sun2014revealing, zhu2014resilience, zio2012analyzing, koc2014structural, kocc2014impact, wang2015network, arianos2009power}.
These studies can have different purposes such as (1) identifying the vulnerable or most influential components of the system in the cascade process by utilizing standard centrality metrics \cite{wei2018novel, wei2018complex, wei2019identification, wang2019modeling, zang2019adjacent, yang2019graph, wei2019electrical, hines2008centrality, wang2011electrical} or defining new centrality metrics \cite{qi2015interaction, qi2018efficient, ju2015simulation, ju2017multi} and (2) identifying the set of components whose upgrade (for instance, by increasing the power flow capacity of transmission lines) or protection can help in  mitigating the risk of cascading failures and large blackouts \cite{ hines2017cascading, zhou2018can, zhou2019markovian, kocc2014impact} or quantifying the performance of the grids after addition of new transmission lines \cite{kocc2014impact, wang2015network}.
To~characterize the latter, some works \cite{wei2019identification, zhu2013risk, sun2014revealing, zhu2014resilience, wang2015network} focus on the response of power grids to attacks and failure scenarios using metrics that quantify the efficiency of the grid before and after the attacks.
Furthermore, to characterize the role of components in the cascade process, some efforts are focused on characterizing the patterns and structures in interaction graphs using community detection approaches \cite{nakarmi2018analyzing, nakarmi2019critical} or tree structures \cite{guo2018failure,guo2017monotonicity}. 
Moreover, the work in \cite{guo2018failure} uses interaction graphs to identify transmission lines that when switched off create partitions that limit the propagation of failures in the power grid.
Some studies \cite{yang2017vulnerability, zhou2019markovian} also use the structures and patterns in interaction graphs to characterize and predict the distribution of cascade sizes.
Similarly, structures in interaction graphs have been used for reliability analysis of zonal patterns \cite{blumsack2009defining} and partitioning into voltage control regions \cite{lagonotte1989structural}.

\subsection{Key Contributions and Review Structure}
The key contributions of this review work are as follows:
\begin{enumerate}[leftmargin=*,labelsep=4.9mm]
    \item A novel classification of methods for constructing interaction graphs into two categories: data-driven and electric distance-based approaches as well as multiple sub-categories as shown in Figure \ref{Fig1:Taxonomy}. 
    \item A comprehensive study and detailed discussion on the techniques used in the construction of interaction graphs. Moreover, key properties and limitations of each type of interaction graph is discussed.  Suggestions on addressing the limitations and possible future directions are presented.
    \item A brief overview of cascading failure analysis in power grids using the constructed interaction graphs are presented.
\end{enumerate}

The organization of the rest of the review is as follows.
In Section \ref{Section2:Definitions}, a brief discussion on the preliminary terms pertinent to this review is presented including: (1) definition and causes of cascading failures, (2) well established cascading failure models in literature, and (3) physical topology-based graphs of power grids.
In Section \ref{Section3:InteractionGraphs}, a comprehensive discussion on the various methods for constructing interaction graphs is provided, and in Section \ref{Section4:ReliabilityAnalysis}, a brief discussion is presented on the reliability analysis and cascading failure studies performed using the interaction graphs constructed in Section \ref{Section3:InteractionGraphs}. Finally, in Section \ref{Summary}, the review is summarized and concluded.

\section{Definitions}
\label{Section2:Definitions}
\vspace{-6pt}

\subsection{Cascading Failures}
\label{Section2:CascadingFailures}
Cascading failures are the leading cause of wide area blackouts \cite{force2004us}.
While large blackouts are infrequent, the power law behavior exhibited by the blackout size distribution 
(e.g., measured in terms of unserved energy, numbers of customers with no service, number of transmission lines tripped) warrants the need to study such events \cite{dobson2007complex}. 
A {\it cascading failure} can be defined as a sequence of interdependent outage events, initiated by few outages  or disturbances \cite{baldick2008initial, vaiman2012risk}. The initiating events can be attributed to various factors such as natural disasters, vegetation disturbances (e.g., tree contact), human errors, software/hardware errors, and so on. In recent years, cyber/physical attacks on power grids, such as the case of the Ukrainian cyber attack of 2015 \cite{liang20162015}, are also precursors to cascading failures.  
After the occurrence of the initiating events, the dependent sequence of outages results from various internal events such as voltage and angular instabilities, line overloads, hidden failures caused due to the misbehavior of protection devices as well as errors related to maintenance, operation, and human factors \cite{vaiman2012risk}. Further, various operating conditions of the power grid, such as the initial loading conditions of the components, also affect the behavior of the overall power grid during cascade~processes.

\subsection{Models of Cascading Failures}
Modeling and studying cascading failures include a diverse field of techniques and approaches. They include topological models, high level statistical models, deterministic and probabilistic models, simulation-based models for analyzing quasi steady  and dynamic behavior of the system, or hybrid models, interdependent models with other systems (e.g., communication systems), and so on (see \cite{guo2017critical}). Many of these methods have been bench-marked and validated as well as cross validated \cite{bialek2016benchmarking, henneaux2018benchmarking}. 
Since this review is focused on graph-based methods, the upcoming sections begin by discussing the physical topology-based graphs of power grids and their limitations in representing interactions among components during cascading failures, which motivates the subsequent discussion of interaction graphs for power grids. 

\subsection{Physical Topology-Based Graphs of Power Grids}
As mentioned in Section \ref{Introduce}, initial graph-based studies of power grids, such as \cite{buldyrev2010catastrophic, crucitti2004model, albert2004structural, holmgren2006using}, were based on the physical topology of the power grid.
In general, a power grid can simply be represented by a graph, $G=(V,E)$, where $V$ represents the set of generator, transmission, substation, or load buses, and $E$ represents the set of power lines \cite{aksoy2019generative}. 
Such a graph shows the physical connectivity among the components of the system. 
Various studies have been performed on the physical topology of the power grids by analyzing their global structural properties \cite{albert2004structural, holmgren2006using}, such as average path length, clustering coefficient, and degree distribution, for analyzing power grids with respect to standard complex networks such as small world, random, and scale-free graphs.
Particularly, in the study in \cite{holmgren2006using}, the average path length and clustering coefficient of real-world power grids were compared to their equivalent random and scale-free network models. However, the study concluded that real-world power grids differed significantly from standard network models as the clustering coefficient and the average path length of real-world grids were significantly greater than that of their complex network model counterparts.
Some studies performed on the physical topology also focused on properties of the electrical connections {\cite{ wang2010electrical} identified using centrality measures such as degree, eigenvector, closeness, and betweenness (for a review and definition of centrality measures refer to \cite{borgatti2006graph})}. 
However, it has also been discussed that physical graphs may be inadequate in representing and capturing the interactions among the components of the power grid \cite{cotilla2012comparing, hines2010topological}, specifically for analyzing cascading failures. 
This limitation is due to the inability of the physical graphs to capture the dynamics of interactions at-distance in cascading failures, for instance, due to Kirchoff's and Ohm's laws.
In recent literature, some studies consider the physical as well as electrical properties of power grids. The focus of such studies is on the generation of synthetic power grid networks that consider the heterogeneity of the components in terms of their operating voltages \cite{halappanavar2015network, aksoy2019generative}. 
In such types of graphs, each vertex is associated with a voltage rating such that transmission lines are represented as edges between vertices of the same voltage level and voltage transformers are represented as edges between vertices with different voltage levels. However, extensive analysis of such graphs for cascading failure scenarios is still an open research problem.

\section{Graph of Interactions}
\label{Section3:InteractionGraphs}
In this section, the methods of modeling power systems by graphs of interactions are reviewed in two distinct categories: data-driven methods and electric distance-based methods. These methods build a graph of interactions for the system, denoted by $G_i=(V_i,E_i)$ in which the set of vertices $V_i$ are the components of the system whose interactions are of interest, such as the set of buses or transmission lines. Further, the set $E_i$ represents the set of interactions/influences among the components, which may be directed, undirected, weighted (representing the strength of interactions or influences), or unweighted depending on the analysis of interest.

\subsection{Data-Driven Methods for Interaction Graphs}
\label{DataDriven}
Various data-driven approaches have been proposed for inferring and modeling interactions among the components of the power grid. These approaches rely on data from simulation or historical outage datasets. 
As the historical datasets are limited, the majority of studies use simulation data.
However, the focus of this review is not on reviewing the mechanism for generating cascade data; for example, from power system simulations. Rather, the focus is on modeling the cascade data into interaction graphs and the subsequent reliability analysis performed on such interaction graphs. The~cascade datasets used in the various methods will be mentioned as necessary when discussing the methods. 
In this paper, five classes of data-driven methods have been identified and reviewed for modeling interaction graphs for studying cascading failures in power grids as shown in Table \ref{TaxonomyData-driven}. Next, each class of method is discussed in detail.

\begin{table*}[t]
{
\small
\caption{Classification of existing studies using the data-driven taxonomy.}
\begin{center}
\begin{tabular}{@{}lllll@{}}
\toprule
\textbf{Category} & \textbf{Subcategory} & \multicolumn{2}{c}{\textbf{Further Subcategory}} & \textbf{Works} \\ 
\midrule
\multirow{7}{*}{\begin{tabular}[l]{@{}l@{}}Data-driven \\ Interaction Graphs\end{tabular}} 
& \multirow{5}{*}{\begin{tabular}[l]{@{}l@{}}Outage Sequence \end{tabular}} 
& \multicolumn{2}{l}{Consecutive Failures} & \begin{tabular}[l]{@{}l@{}}\cite{hines2013dual}, \cite{wei2018novel}, \cite{wei2018complex}, \cite{wei2019identification}, \cite{wang2019modeling}, \cite{zang2019adjacent}, \cite{yang2019graph}, \cite{wei2019electrical}\end{tabular} \\ \cmidrule{3-5} 
 &  & \multicolumn{2}{l}{Generation-based Failures} & \begin{tabular}[l]{@{}l@{}} \cite{qi2015interaction}, \cite{qi2018efficient}, \cite{ju2015simulation}, \cite{ju2017multi}, \cite{chen2019mitigation}, \cite{luo2017identify} \end{tabular} \\ \cmidrule{3-5} 
 & & \multicolumn{2}{l}{Influence-based} & \cite{hines2017cascading}, \cite{zhou2018can}, \cite{nakarmi2018analyzing}, \cite{nakarmi2019critical} \\ \cmidrule{3-5} 
& & \multicolumn{2}{l}{Multiple and Simultaneous Failures} & 
\begin{tabular}[l]{@{}l@{}} \cite{zhou2019markovian}, \cite{wu2007state}, \cite{chang2011performance}, \cite{li2018temporal}, \cite{li2019quantify,li2020state}, \cite{liu2020identifying} \end{tabular} \\ \cmidrule{2-5} 
 & \multicolumn{3}{l}{Risk-graph} & \cite{zhu2013risk}, \cite{sun2014revealing}, \cite{zhu2014resilience}, \cite{kaplunovich2014statistical} \\ \cmidrule{2-5} 
 & \multicolumn{3}{l}{Correlation-based} & \cite{nakarmi2018analyzing}, \cite{nakarmi2019critical}, \cite{yang2017vulnerability} \\ 
 \bottomrule
\end{tabular}
\label{TaxonomyData-driven}
\end{center}
}
\end{table*}

\subsubsection{Interaction Graphs Based on Outage Sequences in Cascading Failures}
\label{OutageSequences}
This class of methods rely on cascading failure data in the form of a sequence of failures in each cascade. For instance, the sequence $l_{5} \rightarrow l_{7} \rightarrow l_{3} \rightarrow l_{6}$ represents an example of sequence of transmission line failures in a cascade scenario, where $l_i$ represents outaged transmission lines and the arrow represents the order in which the lines failed during cascades.
These methods are based on analysis of sequence of failures for extracting interactions and focus on the cause and effect interactions among failure of components. Methods in this category use various techniques and statistics to analyze such data as discussed next.
 
\paragraph{Interaction Graph Based on Consecutive Failures}
\label{ConsecutiveFailuresStatistics}
In this category of outage sequence analysis, only direct consecutive failures in a sequence are used for deriving the interaction links among the components of the system.
In other words, two~components in the system have an interaction link, $e_{i,j} \in E_i$, only if they appear as successive outages in the order $l_{i} \rightarrow l_{j}$ in a cascade scenario in the dataset. 
The order of the outages represents the direction of the links in the interaction graph, e.g., outage sequence $l_{5} \rightarrow l_{7}$  suggests an outgoing link from node $l_5$ to node $l_7$.
Further, if the example sequence $l_{5} \rightarrow l_{7} \rightarrow l_{3} \rightarrow l_{6}$ represents a longer sequence of transmission line failures in a cascade scenario, the following directed interaction edges will belong to graph $G_i$, i.e., \{$e_{5,7}, e_{7,3},$ and $e_{3,6}\}\in E_i$.
The strength of interactions among the components in this case can be characterized using the statistics of occurrences of pairs of successive outages in cascade scenarios in the dataset. For instance, the work in \cite{hines2013dual} assigns weights to the interaction edges by statistical analysis of the number of times that a pair of successive line outages occurs in the cascade dataset. For instance, the weight of the interaction link from node $l_a$ to node $l_b$ can be characterized as $|l_a \rightarrow l_b|/(\text{total number of successive pairs in the overall cascade} \newline \text{dataset})$, where~$|l_a \rightarrow l_b|$ is the number of times failures $l_{a}$ and $l_{b}$ occurred successively in the cascade dataset. These weights can be interpreted as the probability of occurrence of each pair of successive line outages. Examples of studies using this method to develop the power grid's graph of interactions include \cite{hines2013dual, wei2018novel, wei2018complex, wei2019identification, wang2019modeling, zang2019adjacent, yang2019graph, wei2019electrical}, where transmission lines in the system are the vertices $V_i$ of the interaction graph $G_i$.

In the study presented in \cite{wang2011vulnerability}, the sequences of consecutive failures are called {\it fault chains}.
For~creating the dataset of fault chains, in the first step, a single transmission line is tripped as the initiator of cascading failure in the simulation, and in the subsequent steps, the most overloaded component due to power flow re-distributions is considered as the next failure in the overall sequence of consecutive failures.
In the studies in \cite{wei2018novel, wei2018complex, wei2019identification, zang2019adjacent, yang2019graph, wei2019electrical}, for a power grid with $n$ transmission lines, $n$~fault chains are created, and the edges among consecutive failures in each chain are weighted based on power flow changes in a line after the failure.
In the work in \cite{wang2019modeling}, fault chains are created by considering multiple initial failures such that a system with $n$ transmission lines may have more than $n$ fault chains. In addition to power flow re-distributions, the work in \cite{yang2019graph} also considers the temperature evolution process of transmission lines during cascades while constructing fault chains. 
Thus, the fault chains are capable of reflecting thermo-physical effects of transmission lines during cascades.  
Finally, a fault chain graph is developed by combining all fault chains together into a single graph where the vertices are all the components that have failed in the fault chains, and the edges between the vertices exist if the outages have successively occurred in the fault chains. For pairs of outages ($l_i \rightarrow l_j$) that have reoccurred in multiple fault chains, their combined edge weight in the fault chain graph is averaged.

\paragraph{Interaction Graph Using Generation-based Analysis of Failures}
\label{Generation-basedFailures}
The method based on the consecutive failures discussed in Section \ref{ConsecutiveFailuresStatistics} focuses on one to one impact that the outage of a line has on the outage of another line. However, in cascading failures, instead of pair-wise interactions among successive failures, a group of failures may contribute to failures of other components. Therefore, it is important to consider the effects of groups of failures and characterize interactions among the components based on the effects among groups of components. The works presented in \cite{qi2015interaction, ju2015simulation, ju2017multi, qi2018efficient, chen2019mitigation} define such groups as {\it generation} of failures within a cascade process, which are failures that occur within short temporal distance of each other.
In these works, the sequence of failures in the cascade are divided into sequence of generations, and the failure induced cause and effect relationships are considered between consecutive generations. Specifically, outages occurring in generation $m+1$ are assumed to be caused by outages in generation $m$.

The interactions based on successive generations are defined in different ways in the literature. For instance, the authors in \cite{hines2017cascading} assume that all components in generation $m$ have interactions with all components in generation $m+1$, i.e., if generation $m$ has $n_1$ number of components and generation $m+1$ has $n_2$ number of components, then the number of interactions between generation $m$ and $m+1$ will be $n_1 \times n_2$. However, some studies argue that considering all possible pairs of interactions among components of two consecutive generations overestimates the interactions among components \cite{qi2015interaction, ju2015simulation, ju2017multi, luo2017identify, qi2018efficient, chen2019mitigation}. Specifically, all line outages in one generation may not be the cause of a line outage in the next generation. Therefore, in the works presented in  \cite{qi2015interaction, ju2015simulation}, the cause of failure of a line $k$ in generation $m+1$ is considered to be due to the failure of a line in generation $m$ with the maximum influence value on the line $k$. The influence value for component $j$ in generation $m$ is defined as the number of times that the component $j$ has failed in generation $m$ before the failure of line $k$ in the successive generation $m+1$ in the cascade dataset.
For cases where two or more lines in generation $m$ have the same maximum influence values on line $k$ in generation $m+1$, all such components are assumed to interact with line $k$.
In the works discussed so far in this section, the interaction among component $j$ in generation $m$ and component $k$ in generation $m+1$ will be represented by a directed link $e_{j,k}$. 

While the works in \cite{qi2015interaction, ju2015simulation} limit the interactions by only considering the maximum influence values in current generation as the probable cause of component failures in the next generation, the~work in \cite{qi2018efficient} gives an estimate of the interactions between successive generations using the expectation maximization (EM) algorithm. Initially, all failed components in generation $m$ are assumed to be the causes of failure of all components in generation $m+1$.
However, the actual components in generation $m$ (hidden variables) that cause failure of components in generation $m+1$ are found using the iterative process of updating the probabilities of failures. Thus, after the iterative update process is completed, some probabilities of failures between components may be zero, which removes the overestimated interactions of the initial assumption.

The weight of the interaction links can also be defined in various ways. For instance, the weight of the link can be defined as the ratio of number of times that the pair of components appeared in two successive generations over the total number of times that component $k$ has appeared in the dataset. This weight can be interpreted as the likelihood of failure of component $k$ in the next generation given the failure of component $j$ in the current generation. The work in \cite{ju2017multi} also generates the graph of interactions based on the maximum influences among generations in cascades, similar to the method used in \cite{qi2015interaction, ju2015simulation}. However, the graph of interactions has two additional layers: one layer to capture the weight of interactions based on the amount of load shed that has occurred after the failures in generation $m$ (therefore, the dataset requires additional information about the amount of load shed during the cascade process) and the other layer to capture the weights of interactions based on the electric distance between transmission lines during the cascade process. We will present the detailed discussion of the electric distance-based interactions in Section \ref{ImpedanceSection}. The study in \cite{luo2017identify} considers both statistical properties as well as the amount of load shed that has occurred between successive generations to assign interaction link weights. However, the study in \cite{luo2017identify} identified islands formed in the power grids during outages and then selectively assigned links between components of successive generations only if the generations were located in islands that were direct consequences of one~another.

\paragraph{Influence-Based Interaction Graph}
\label{InfluenceGraphs}
In this method, the interactions among the components are derived based on successive generations in cascades; however, the weights of the interactions are characterized based on the influence model and the branching process probabilistic framework. The influence model is a networked Markov chain framework, originally introduced in \cite{asavathiratham2001influence} and was first applied to cascade dataset in the work presented in \cite{roy2001network}. 
This survey reviews the studies that use the influence model in the context of power grids to develop the graph of interactions. In these studies, the transmission lines in the system are considered as the vertices $V_i$ of the interaction graph $G_i$ and the influences/interactions between the lines as the edges $E_i$.

In \cite{hines2017cascading}, authors consider interaction links among all pairs of lines in two successive generations in a cascade using the influence model. The weights of the directed links are derived in two steps. In~the first step, a branching process approach is used in which each component can produce a random number of outages in the next generation. The number of induced outages by each component is assumed to have a Poisson distribution based on the branching process model. Parameter $\lambda_i$ specifies the propagation rate (mean number of outages) in generation $m+1$ for the outage of component $i$ in generation $m$. In other words, this step defines the impact of components on the process of cascade by describing how many failures their failure can generate \cite{dobson2012estimating}.
In the second step, it is assumed that given that component $i$ causes $k$ outages in the next generation, some components are more likely to outage than others. Therefore, they calculate the conditional probability $g(j|i)$, which is the probability of component $j$ failing in generation $m+1$, given the failure of component $i$ in generation $m$. If only $g(j|i)$ values based on the statistical analysis of data are considered, then the probability of failure of component $j$ given component $i$ failure will be known; however, the expected number of failures from failure of component $i$ is not known. Hence, both steps are important in characterizing the influences among components.

The final step consists of combining the information from the first and second steps into a single influence matrix \textbf{H} (representing the links of graph of interactions and their weights). The elements of the matrix are defined based on the conditional probability that a particular component $j$ fails in the next generation $m+1$, given that component $i$ has failed in generation $m$ and that generation $m+1$ includes exactly $k$ failures. This probability can be defined as $P(j|{i,k})=1-(1-g[j|i])^k$.
Then, the~conditional probability $h_{i,j,m}$ that component $j$ fails in generation $m+1$, given that component $i$ failed in generation $m$, over all possible values of $k$ represents the actual elements of \textbf{H} and is found by multiplying $P(j|{i,k})$ with the probability of $k$ failures occurring as follows:

\begin{equation}
    h_{i,j,m} = \sum \limits_{k=0}^{\infty} (1 - (1 - g[j|i]^{k})) \frac{\lambda_{i,m}^{k}}{k!} e^{-\lambda_{i,m}}.
\end{equation}

Based on the influence graph, cascading failures can start with a line outage at a node of the graph and propagate probabilistically along the directed links in the graph. 
Examples of other works, that have used the influence-based approach to derive the graph of interactions for power grids include~\cite{zhou2018can,nakarmi2018analyzing, nakarmi2019critical}.

\paragraph{Interaction Graph of Multiple and Simultaneous Failures}
\label{MarkovianInteractionGraph}
This class of methods also use the sequence of failures to model interactions among components of the system; however, they consider the interactions among multiple simultaneous failures.

Specifically, in the study presented in \cite{zhou2019markovian}, a Markovian graph was developed with the goal of addressing
the problem of capturing the effect of multiple simultaneous outages within generations on the characterization of the interactions among the components of the successive generations in a cascade.
In this case, the nodes of the graph represent the states of the Markov chain defined as the set of line outages in a generation of the cascade, and the links represent the transition among the states (i.e., interactions between successive generations of outages). Hence, each node in the graph may represent the outage of a single line or multiple lines. 
Markovian interaction graphs differ from generation-based and influence-based interaction graphs as edges are the interactions between successive generations of sets of line outages instead of the individual interactions between line outages in successive generations. Markovian interaction graphs also consider a node with a null state, which represents the state where the cascade stops. This state occurs at the end of all cascade scenarios.
The transition probabilities among the states (i.e., the weight of the links) from state $i$ to state $j$ can be estimated by counting the number of consecutive states in which state $i$ and state $j$ occur in all the cascades and dividing by the number of occurrences of state $i$.

Other studies that consider the interaction among multiple failures at the same time are presented in \cite{wu2007state, chang2011performance}. In these works, the state of each component in the power grid, including buses and transmission lines/transformers, is represented by $0$ for a failed condition and $1$ for a working condition. Then, states of all components in the power grid are aggregated and represented by a vector with size $n + l$, where $n$ is the number of buses and $l$ is the number of transmission lines/transformers. A single state vector can be regarded as a node in the interaction graph, and there are exactly $2^{n+l}$ number of nodes in the graph, representing all possible states. Initially, all components in the power grid are in working condition such that the initial state consists of a state vector $i$ represented by all ones. After a component failure occurs, the state of the component changes to $0$, and the corresponding entry of the component in the initial state vector $i$ is also updated, and thus, a new state vector $j$ is formed. This transition is represented in the interaction graph by a directed link from the initial state vector $i$ to the updated state vector $j$. This process is continued for all sequences of failures in the cascade dataset. Finally, some nodes in the state transition graph may be highly connected whereas some nodes may be isolated. Note the obtained interaction graph is directed but unweighted.

Other examples of methods that consider interactions of multiple failures at the same time include  \cite{li2018temporal, li2019quantify, li2020state, liu2020identifying}. In these works, fault chains are used to construct a {\it state failure} network. Each fault chain is also associated with its final load loss value. Note that the definition of a fault chain has already been discussed in Section \ref{ConsecutiveFailuresStatistics}. 
Similar to the studies in \cite{wu2007state, chang2011performance}, each component is represented by binary values $0$ or $1$ for working and failed conditions, respectively. The working/failed conditions for all components in the system are aggregated to form a state.
Each state is an $n$-dimension vector, where $n$ is the number of transmission lines in the power grid. However, in contrast to the studies in \cite{wu2007state, chang2011performance}, where all possible states, i.e., $2^n$, are considered as nodes in the interaction graph, states are enumerated from the fault chain data. For example, for a fault chain $\{l_{5} \rightarrow l_{7} \rightarrow l_{3} \rightarrow l_{6}\} \rightarrow load_{loss}$, four states are used for representing the four line faults in the chain, and a single state is used to represent the total load loss associated with the fault chain. All such fault chains in the cascade dataset are combined together into the state failure network. Nodes in the network represent the states enumerated from the complete dataset, and links in the network represent the failed component occurring immediately after a particular state. Thus, the links can be regarded as records of component failures that occur after a state. 
Next, nodes and links in the network are assigned weights by back-propagation of the final load losses based on the probability of occurrence of the links. Thus, each node (state) and each link (failed component) is associated with an expected load loss after their occurrence.

\subsubsection{Risk Graphs for Interaction Graph}
\label{RiskGraphs}
The work presented in \cite{zhu2013risk} introduces the {\it risk-based} interaction graphs, which  describe the interactions or relationships among the nodes (i.e., buses/substations) of the power grid based on the effects of their simultaneous failures in causing damage in the system. This graph is not solely focused on analysis of interactions among components during cascading failures. Instead, it is focused on the vulnerability analysis of the power grid, and the effect of failures is assessed using metrics such as {\it net-ability}, which measures the effectiveness of a power grid subjected to failures, based on power system attributes including power injection limitation and impedance among the components.

Construction of risk graphs is done in two steps. The first step includes generating and tracking the sets of strongest node combinations whose simultaneous failures have significant effects on the power grid. Identification of such sets of strong node combinations can be done by reducing the search space or exhaustive search methods \cite{zhu2013risk, sun2014revealing}.
Reduced search space strategy is the preferred method for computational purposes. For instance, in \cite{sun2014revealing}, the search works as follows: given $m$-node combination of components, which causes damage in the network, $m+k$-node combination (where $k$ represents additional components) should cause an even greater damage. 

In the second step, these sets of strong node combinations are used to form the {\it risk graphs}.
If a node appears at least once in the sets of strong node combinations, then the node becomes a vertex of the risk graph. Links among nodes in the risk graph exist if they appear in the same set of strong node combinations. Both nodes and links in the risk graph are weighted based on the frequency of their appearance in the sets of strong node combinations. This approach results in a weighted but undirected node risk graph, where higher weight values on the links suggest stronger node combinations. Node risk graphs are dependent on the system parameters such as ratio of capacity to the initial load of the nodes in the system. To remove dependencies on system parameters, node risk graphs can be constructed for multiple parameter values and combined together to form the node integrated risk graph using the risk graph additivity property \cite{zhu2013risk, sun2014revealing}.
The aforementioned risk graph can also be extended to a directed risk graph, where the removal of components in a specific order in strong node combinations are considered. The study in \cite{zhu2014resilience} constructs the directed node risk graphs and the directed node integrated risk graph with the same concept as its undirected counterpart in the studies in \cite{zhu2013risk} and \cite{sun2014revealing}.

Another similar concept to risk graph is the {\it double contingency graph} introduced in \cite{kaplunovich2014statistical}.
While~$m$ contingency combinations of attack scenarios for the power grid was studied in the risk graphs, many~methods focus on N-2 contingency analysis as the power grid is considered to be N-1 protected \cite{nercReliabilityStandard2018}.
In the double contingency graph, the vertices of the graph are the transmission lines, and the links between vertices show pairs of transmission lines whose simultaneous failure as initial triggers can affect the reliability of the system by, for instance, violating the thermal constraint rules in the power grid. Similar to the risk graph, double contingency graph only considers combinations of initial triggers and lacks information about the components, which will be affected due to the outage of the initial triggers. Therefore, the work in \cite{kaplunovich2014statistical} uses a combination of the double contingency graph with influence graph for reliability analysis of the power grid.

\subsubsection{Correlation-Based Interaction Graph}
\label{CorrelationGraphs}
The work in \cite{yang2017vulnerability} presents a graph of interactions for power grids based on the correlation among the failures of the components. In the correlation-based interaction graph in \cite{yang2017vulnerability}, vertices represent the transmission lines, and the edges represent the pairwise correlation between line failures in the cascade dataset. The correlation dependence between failures are captured in the correlation matrix, whose $ij$th elements are positive Pearson correlation coefficient between the failure statuses of components $i$ and $j$ in the cascade dataset. The resulting correlation matrix is symmetric and can be interpreted as an undirected and weighted interaction graph, where the nodes are the failed lines, the edges are the interactions between the lines, and the weights are the correlation values among the components.
Similarly, the studies in \cite{nakarmi2018analyzing} and \cite{nakarmi2019critical} also construct correlation-based interactions graphs from simulated cascade dataset consisting of sequences of transmission line failures.

\begin{table*}[t]
{
\begin{center}
\small
\caption{Key properties and limitations of the methods for building data-driven interaction graphs.} \label{Data-drivenStrengthsandLimitations}
\begin{tabular}{|l|l|l|l|l|}
\hline
\multicolumn{2}{|l|}{Category} & Key Properties & \multicolumn{2}{l|}{Limitations} \\ \hline
\multicolumn{2}{|l|}{\begin{tabular}[c]{@{}l@{}}Outage \\ Sequence: \\ Consecutive \\ Failures\end{tabular}} & \begin{tabular}[c]{@{}l@{}}\tabitem Simple derivation of interactions based on \\ consecutive order of failures, resulting in directed graphs.\\ \\ \tabitem Weights of interactions are assigned using various \\ methods, e.g., statistical analysis or physical/electrical \\ properties of the system.\end{tabular} & \multicolumn{2}{l|}{\begin{tabular}[c]{@{}l@{}}\tabitem Only considers direct one-to-one \\ consecutive interactions in cascades.\\ \\ \tabitem Does not consider interactions \\ between groups of components.\end{tabular}} \\ \hline
\multicolumn{2}{|l|}{\begin{tabular}[c]{@{}l@{}}Outage \\ Sequence: \\ Generation-\\ based \\ Failures\end{tabular}} & \begin{tabular}[c]{@{}l@{}}\tabitem Interactions are considered between groups of \\ components (based on the concept of generations) and are \\ used to define one-to-one interaction links between \\ components of consecutive generations.\\ \\ \tabitem Considers order of failures/interactions, resulting \\ in directed graphs.\end{tabular} & \multicolumn{2}{l|}{\begin{tabular}[c]{@{}l@{}}\tabitem Interactions between groups of \\ components of two consecutive generations \\ characterized using various methods \\ (may overestimate/underestimate \\ the interactions).\end{tabular}} \\ \hline
\multicolumn{2}{|l|}{\begin{tabular}[c]{@{}l@{}}Outage \\ Sequence: \\ Influence-\\ based\end{tabular}} & \begin{tabular}[c]{@{}l@{}}\tabitem Generation-based interactions are considered to \\ define one-to-one interaction links between components \\ based on the influence framework.\\ \\ \tabitem Considers order of failures/interactions, resulting \\ in directed graphs.\\ \\ \tabitem Considers propagation rate of line outages as \\ well as their probability of causing further outages \\ through influence framework.\\ \\ \tabitem Influence-based framework provides mathematical \\ tractability for certain analysis of the dynamics of failures.\end{tabular} & \multicolumn{2}{l|}{\begin{tabular}[c]{@{}l@{}}\tabitem Interaction links among all pairs of \\ components of two consecutive generations \\ may overestimate the interactions.\end{tabular}} \\ \hline
\multicolumn{2}{|l|}{\begin{tabular}[c]{@{}l@{}}Outage \\ Sequence: \\ Multiple and\\ Simultaneous\\ Failures\end{tabular}} & \begin{tabular}[c]{@{}l@{}}\tabitem Characterizes simultaneous interactions between \\ groups of components instead of only one-to-one \\ interactions between pairs of individual components.\\ \\ \tabitem Considers order of failures/interactions, resulting \\ in directed graphs.\\ \\ \tabitem Considers group interactions, eliminating the \\ challenge of overestimation or underestimation of \\ pair-wise interactions between individual components.\end{tabular} & \multicolumn{2}{l|}{\begin{tabular}[c]{@{}l@{}}\tabitem The number of nodes in the resulted \\ graphs (to capture possible states of group \\ interactions) will be large.\end{tabular}} \\ \hline
\multicolumn{2}{|l|}{Risk-graph} & \begin{tabular}[c]{@{}l@{}}\tabitem Characterizes the interactions based on targeted \\ failures of the components, resulting in directed graphs.\\ \\ \tabitem Requires small dataset for small number of \\ targeted failures.\end{tabular} & \multicolumn{2}{l|}{\begin{tabular}[c]{@{}l@{}}\tabitem Interactions between groups of \\ components are not considered.\\ \tabitem Reduced performance when \\ number of targeted failures increase.\end{tabular}} \\ \hline
\multicolumn{2}{|l|}{Correlation-based} & \begin{tabular}[c]{@{}l@{}}\tabitem Uses simple statistical correlation measure to \\ construct the graph.\end{tabular} & \multicolumn{2}{l|}{\begin{tabular}[c]{@{}l@{}}\tabitem Does not consider the order of \\ failures in interactions.\\ \tabitem Does not consider group interactions.\\ \tabitem Resulted graphs are usually dense.\end{tabular}} \\ \hline
\end{tabular}
\end{center}
}
\end{table*}

\subsubsection{Comparison of Data-Driven Methods for Constructing Interaction Graphs}

In the past decade, cascading failures have been actively modeled and analyzed using data-driven interaction graphs.
These graphs can be considered as abstract models of power grids for studying cascading failures; as the physical details of the system such as generation, load consumption, line~power flows, capacity constraints, etc., are not explicitly considered; instead, they are implicitly captured through the cascade data. 
This section provides a brief comparison among the data-driven interaction graphs and discusses their key properties and limitations (summarized in Table \ref{Data-drivenStrengthsandLimitations}).

One key difference among the data-driven methods for constructing interaction graphs is the type of data that they need to build the graph. For instance, influence-based methods can be applied to both simulation data as well as historical data. However, for the consecutive failure-based methods, such as fault-chains, as well as risk graphs, the sequence of failures needs to be generated by targeted failure of components to create a comprehensive list of failures to build the interaction graph. Moreover, as~the combinations of the initial targeted failures increase, the computational complexity of generating the list of failures and building the risk graphs also increases.

Another difference among the data-driven methods is their ability in considering group interactions; e.g., the consecutive failures-based methods can only consider one-to-one interactions based on direct consecutive orders of failures. However, generation-based and influence-based methods can consider the influence of a group of components on another group of components (using the concept of generations) and use it to define one-to-one interactions among pairs of components. Multiple and simultaneous failures interaction methods not only consider the influence of a group of components on another group but also allow for interaction links among groups of components instead of individual pairs of components. 
The data-driven methods for constructing interaction graphs result in directed graphs except for the correlation-based method. The correlation-based method cannot capture the order and direction of interactions due to symmetric properties of the correlation measure. Another challenge with the correlation-based method is that the method generally results in dense graphs, with many links showing small correlations among the components, which may need to be thresholded or applied with other techniques to be able to focus on more significant correlations.   
In general, one of the key limitations of the data-driven methods is the heavy dependency of the derived graphs and inferred interactions on the dataset and the applied method, which may lead to overestimation/underestimation of the interactions.
As shown in \cite{nakarmi2018analyzing, nakarmi2019critical}, interaction graphs derived from different outage data are different and can lead to different conclusions; e.g., the operating setting of the power grid (such as its loading level) will affect the cascade process and the cascade dataset \cite{OperatingSetting} and, consequently, lead to different interaction graphs. Using a single interaction graph to model power grids and capture all possible operating setting conditions that can affect the system is still a challenge to address. 
There is also a need to test the interaction graphs to historical cascade data \cite{vaiman2012risk}. Historical data shows the heavy-tailed distribution of blackout sizes and can be used to benchmark the data-driven interaction graphs \cite{vaiman2012risk}. Similarly, cross validating the various techniques with each other may also provide insights into the most useful techniques in extracting interactions from cascade data.

\subsection{Electric Distance-Based Interaction Graphs}
While there is abundant literature focused on data-driven interaction graphs, various methods have been proposed for modeling interactions among the components using the dynamics of power flow as well as the physical/electrical properties of the system and components. 
In this review, such~methods are called electric distance-based methods.

In a power grid, electricity does not flow through the shortest path between two nodes $i$ and $j$. 
Instead, it can flow through parallel paths between nodes $i$ and $j$ based on the physical properties of the system and its components as well as the physics of electricity (i.e., Ohm's law). 
Thus, the~electrical interactions between the components may extend beyond the physical topology and the direct connections in the power grid. The concept of electric distance was first introduced by Lagonotte et al. \cite{lagonotte1989structural} in 1989 as a measure of coupling between buses in the power system and was based on sensitivities in the power system due to changes in voltage magnitudes. There are various methods in the literature for characterizing the electrical distances between components. These methods can use {\it distribution factors} in power grids including {\it power transfer distribution factor} (PTDF) \cite{ejebe2000fast}, which indicates the change in real power on transmission lines due to changes in real power injection at different nodes of the system, and {\it line outage distribution factor} (LODF) \cite{guo2009direct}, which measures the changes in the power flow of transmission lines due to the outage of another line. 
Next, the electric distance-based methods for constructing interaction graphs are discussed in two categories: outage condition-based and non-outage condition-based as shown in Table \ref{TaxonomyElectricDistance-based}.

\subsubsection{Outage Condition-Based Interaction Graph}
This class of methods for characterizing electric distance-based interaction graphs are focused on interactions among the components of the power grid during outage conditions. 
For instance, in~the study presented in \cite{guo2018failure}, interactions among the components as well as their weights are derived using the changes in the power flows in transmission lines during outage conditions.
Thus, the outage induced interaction graph $G_i$ consists of the set of vertices $V_i$ that represent the transmission lines and the set of edges $E_i$ that represent the impact of outage of one line on another.
This impact is characterized using LODF \cite{guo2009direct}, where LODF  for line $e_{i,j} \in E_i$ is calculated based on the ratio of the impact of outage of line $i$ on line $j$ based on the reactance of all possible spanning tree paths between the lines, over the impact of outage of line $i$ on line $j$ using the reactance of all alternative spanning tree paths where the power can flow (i.e., excluding the spanning tree path of line $i$) \cite{guo2017monotonicity}.  

However, during cascading failures, the impact of a failed line on the remaining lines is not limited to changes in power flows. In a power grid, if two or more lines share a bus, outage of one line may expose the remaining lines (connected through the same bus) to incorrect tripping due to malfunctioning of the protection relays. The exposed lines are prone to failure and an increase in power flow in the exposed lines exacerbates their tripping probability causing further outages. Such failures are known as hidden failures. 
In the studies in \cite{ma2017speeding, ma2017fast, ma2018fast}, vertices $V_i$ represent transmission lines as well as a hidden failure state, and edges $E_i$ represent inter-line interactions as well as interactions between lines and the hidden failure state.
Thus, the interaction graph $G_i$ will have $n+1$ nodes where $n$ is the number of transmission lines, and the extra one node represents the hidden failure state.
The~hidden failure node has bidirectional links from itself to every other node in the power grid. However, the~hidden failure node does not have any influence on itself.
The inter-line interaction $e_{i,j} \in E_i$ shows the increase of power flow in line $j$ due to outage of line $i$. 
The interaction from the hidden failure node to a line $i$ reflects the likelihood of failure of line $i$ due to the failure of other nodes when the power flow in line $i$ exceeds a flow limit margin. Interaction from line $i$ to the hidden failure node reflects the average tripping probability of all the other lines due to the outage of line $i$.
This~interaction can be regarded as the aggregated influence that the outage of line $i$ has on the tripping probability of all the other remaining lines.

\begin{table*}[t]
{
\small
\caption{Classification of existing studies using the electric distance-based taxonomy.}
\begin{center}
\begin{tabular}{@{}lllll@{}}
\toprule
\textbf{Category} & \textbf{Subcategory} & \multicolumn{2}{c}{\textbf{Further Subcategory}} & \textbf{Works} \\ 
\midrule
\multirow{3}{*}{\begin{tabular}[l]{@{}l@{}}Electric Distance-based \\ Interaction Graphs\end{tabular}} 
& \multicolumn{3}{l}{Outage Condition-based} &
\begin{tabular}[l]{@{}l@{}}\cite{guo2017monotonicity}, \cite{guo2018failure}, \cite{ma2018fast}, \cite{ma2017speeding}, \cite{ma2017fast}\end{tabular} \\ \cmidrule{2-5}
& \multirow{2}{*}{Non-outage Condition-based} & \multicolumn{2}{c}{Impedance-based}  
& \begin{tabular}[l]{@{}l@{}} \cite{ju2017multi}, \cite{zio2012analyzing}, \cite{poudel2018electrical}, \cite{hines2008centrality}, \cite{blumsack2009defining}, \cite{cuffe2015visualizing}, \cite{bompard2009analysis}, \\ \cite{wang2011electrical}, \cite{bompard2012extended}, \cite{arianos2009power}, \cite{koc2014structural}, \cite{kocc2014impact}, \cite{wang2015network} \end{tabular}\\ \cmidrule{3-5}
&  &  \multicolumn{2}{c}{Jacobian} & \begin{tabular}[l]{@{}l@{}} \cite{cuffe2015visualizing}, \cite{caro2015voltage}, \cite{caro2015centrality}, \cite{cotilla2012comparing}, \cite{cotilla2013multi} \end{tabular} \\
\bottomrule
\end{tabular}
\label{TaxonomyElectricDistance-based}
\end{center}
}
\end{table*}

\subsubsection{Non-Outage Condition-Based Interaction Graphs}
\label{ImpedanceSection}
This class of methods for constructing the electric distance-based interaction graphs is focused on interactions among the components of the power grid during normal operating conditions. For~instance, measures to capture the characteristics of the power flow paths between components, such~as impedance of the transmission lines, can be used to form electric distance-based interaction graphs. Moreover, power system sensitivities based on PTDF and the ones showing changes in voltage magnitudes and voltage phase angles, derived from Jacobian matrices during normal conditions, can also be used to form electric distance-based interaction graphs.~{In this paper, non-outage condition-based interaction graphs are broadly categorized into two categories: impedance-based and Jacobian. Next, both categories are discussed in detail.}

\paragraph{Impedance-Based Interaction Graph}
Impedance-based electric distance interaction graphs $G_i$ consist of vertices $V_i$ that represent buses and edges $E_i$ that represent electrical interactions between pairs of buses weighted by their corresponding impedance-based electrical distances. 
Inverse admittance matrix, more commonly known as the impedance matrix \textbf{Z}, is one of the simplest forms of representing electrical interactions between pairs of buses in the system and is found by inverting the system admittance matrix \textbf{Y}, i.e., 
\begin{math}
\mathbf{Z = Y^{-1}}.
\end{math}
Matrix \textbf{Z} shows the relationship between the nodal bus voltage vector and the nodal current injection vector. However, unlike matrix \textbf{Y}, which is sparse, impedance matrix \textbf{Z} is non-sparse as it represents the changes in nodal voltage throughout the system due to a single nodal current injection between a pair of nodes in the system.
Therefore, edges in the impedance-based interaction graph are the connections between the elements in the \textbf{Z} matrix with weights between buses $i$ and $j$ corresponding to their absolute value of the impedance, i.e., $|Z_{ij}|$ \cite{hines2008centrality, blumsack2009defining, zio2012analyzing, cuffe2015visualizing, poudel2018electrical}.
Smaller magnitudes of impedance represent shorter electric distance between buses.
Note that the individual elements $Z_{ij}$ in matrix $Z$ are complex valued.
The studies in \cite{bompard2009analysis, wang2011electrical, bompard2012extended} also adopt the concept of representing electrical distances between buses using the Thevenin equivalent impedance between buses but apply the condition that power only flows from generator buses to load buses such that impedance values between generator buses and load buses suggest edges in the interaction graph. 
Similarly, in the study in \cite{arianos2009power}, in addition to the impedance between pairs of generator and load buses in the system, the~power flows through the lines along the path between the pairs of buses is considered. Therefore, electrical connections between generator and load buses $i$ and $j$ along path $k$, are weighted by the impedance between buses $i$ and $j$ as well as the PTDF of the lines along the path $k$ of power flows between the buses. In the studies in \cite{bompard2009analysis, wang2011electrical, bompard2012extended, arianos2009power}, matrix $Z$ has dimension $Z_G \times Z_L$, where $Z_G$ and $Z_L$ are the number of generator and load buses, respectively. Thus, matrix $Z$ may be asymmetric, depending on the number of generators and load as well as the entry of the elements suggesting that the interaction graph is directed.

Note that studies in \cite{hines2008centrality, blumsack2009defining, zio2012analyzing, cuffe2015visualizing, poudel2018electrical, bompard2009analysis, bompard2012extended, wang2011electrical, arianos2009power} are not focused on cascading failures, but their concept of formulating impedance-based electrical distances can be extended for studying cascade processes. For example, in the study in \cite{ju2017multi}, transmission lines are considered as the nodes of the interaction graph, and thus, impedance-based electric distances between pairs of transmission lines are assigned as weights of the interaction links. To find the electric distance between transmission lines $i$ and $j$, where line $i$ connects bus $i_s$ to $i_d$ and line $j$ connects bus $j_s$ to $j_d$,
the minimum of the four possible Thevenin equivalent impedances between the pairs of buses, i.e., $Z_{i_sj_s}$, $Z_{i_sj_d}$, $Z_{i_dj_s}$, and $Z_{i_dj_d}$ is taken. This interaction graph is also the third layer of the multi-layered interaction graph discussed in Section \ref{Generation-basedFailures} used for modeling cascading failures.

Cascading failures can also be studied by representing interactions between pairs of nodes by effective resistances between the nodes.
Effective resistance, $R_{ij}$, between nodes $i$ and $j$, also known as {\it Klein resistance distance} \cite{klein1993resistance}, is the equivalent resistance of all parallel paths between the nodes. 
It was initially introduced in the study in \cite{klein1993resistance} as a measure of distance in graph theory, and in the context of power systems, it shows the potential difference between nodes $i$ and $j$ due to unit current injection at node $i$ and withdrawal at node $j$.
While impedance-based electrical distances between nodes account for non-linear approximations of power flow, effective resistances only consider linear approximations of power flows in the grid. 
For cascading failure analysis, as the impedance of a transmission line in a high voltage transmission network is dominated by the imaginary part of impedance, i.e., reactance, the effective resistance between nodes can be formulated in terms of their reactance.
Thus, in the studies in \cite{koc2014structural, kocc2014impact, wang2015network}, effective resistance between nodes $i$ and $j$ is found as 
\begin{math}
        R_{ij} = {Q_{ii}} - {2Q_{ij}} + {Q_{jj}}
\end{math}
where, $Q_{ij}$ is the row $i$ and column $j$ element of $\mathbf{Q^{+}}$, which is the Penrose pseudo-inverse of the Laplacian matrix \textbf{Q}. Matrix \textbf{Q} is defined as the difference between the weighted diagonal degree matrix and the weighted adjacency matrix derived from the physical topology and shows the relationship between the buses and transmission lines in the grid. 
In the studies in \cite{koc2014structural, kocc2014impact, wang2015network}, the weights of the edges in the physical topology required for finding the weighted diagonal degree matrix and adjacency matrix are the susceptance (i.e., the imaginary part of impedance) values between the nodes. Note that matrix \textbf{Q} constructed using susceptance values is equivalent to matrix \textbf{Y}.
Thus, edges $E_i$ in the effective resistance interaction graph reflect the electrical connections between the buses with weights between nodes $i$ and $j$ being the corresponding $R_{ij}$ values.

\paragraph{Jacobian Interaction Graph}

Electric distance-based interaction graphs can also be constructed using the sensitivity matrix of the power grid during normal operating conditions. In such interaction graph $G_i$, vertices $V_i$ represent the buses, and the edges $E_i$ represent the electrical interactions in terms of sensitivities between the buses.
These sensitivities can be found using the Jacobian matrix, which is obtained during Newton--Raphson-based load flow computation. Jacobian sensitivity matrix \textbf{J} shows the effect of complex power injection at a bus on the voltage magnitude and voltage phase angles of other buses. It consists of four sub-matrices: matrix $\mathbf{J_{P\boldsymbol{\theta}}}$, which shows the relationship between nodal active power injections and voltage phase angle changes; matrix $\mathbf{J_{PV}}$, which shows the relationship between nodal active power injections and voltage magnitude changes; matrix $\mathbf{J_{Q\boldsymbol{\theta}}}$, which shows the relationship between nodal reactive power injections and voltage phase angle changes; and matrix $\mathbf{J_{QV}}$, which shows the relationship between nodal reactive power injections and voltage magnitude changes. 
The inverse of any of these Jacobian sub-matrices, denoted as $\mathbf{J^{-1}}$, can be used to find the sensitivity matrix by using the Klein resistance distance \cite{klein1993resistance}, whose individual element is calculated as 
\begin{math}
    x_{ii} + x_{jj} - x_{ij} - x_{ji},
\end{math}
where $x_{ij}$ represents the element in row $i$ and column $j$ of the inverted Jacobian sub-matrix $\mathbf{J^{-1}}$ in consideration. 
In the study in \cite{cuffe2015visualizing}, all of the four Jacobian sub-matrices are applied to Klein resistance distance to form sensitivity matrices for the purpose of visualizing power grids, which in turn can be used to form interaction graphs whose edges $E_i$ represent the electrical interactions between the components of the sensitivity matrices weighted by their corresponding elements.

However, the literature has revealed that most studies are focused on two of the four Jacobian sub-matrices, i.e., matrix $\mathbf{J_{P\boldsymbol{\theta}}}$ and matrix $\mathbf{J_{QV}}$. The remaining sub-matrices $\mathbf{J_{Q\boldsymbol{\theta}}}$ and matrix $\mathbf{J_{PV}}$ are not used in the literature due to un-intuitive interpretations. The seminal work of electrical distances by Lagonotte et al. in \cite{lagonotte1989structural} focuses on using the Jacobian sub-matrix ($\mathbf{J_{QV}}$), also known as the voltage sensitivity matrix $\boldsymbol{\partial} \mathbf{V}/ \boldsymbol{\partial} \mathbf{Q}$, to find the electric distance between buses. Similarly, the study in \cite{caro2015voltage} also uses the voltage sensitivity matrix. 
In both studies, the matrix of maximum attenuations is found, which consists of columns of voltage sensitivity matrix divided by the diagonal values. Finally, electrical interactions $e_{i,j} \in E_i$ between buses $i$ and $j$ weighted by their electric distance is derived as the logarithm of the individual elements of the attenuation matrix. 
The study in \cite{caro2015centrality} also uses the voltage sensitivity matrix to find electric distance between buses, but instead of finding matrix of attenuations, the study applies the sensitivity matrix to Klein resistance distance formulation. Note that studies in \cite{caro2015centrality, caro2015voltage} analyze risk of cascading failures by studying the voltage collapse phenomenon, which is a sequential process during which large parts of the power grid may suffer due to low voltages \cite{kundur2004definition}.
Similarly, the studies in \cite{cotilla2012comparing} and \cite{cotilla2013multi} apply the Jacobian sub-matrix  $\mathbf{J_{P\boldsymbol{\theta}}}$ or the sensitivity matrix $\mathbf{\boldsymbol{\partial} P/\boldsymbol{\partial} \boldsymbol{\theta}}$ to Klein resistance distance and find electrical distances between buses. However, the studies \cite{cotilla2012comparing, cotilla2013multi} are not focused on cascading failure analysis but are included in this review as their electric distance-based interaction graphs can have potential implications for analyzing cascading failures.

\subsubsection{Key Properties and Limitations of Electric Distance-Based Interaction Graphs}
In this section, key properties and limitations of the electric distance-based interaction graphs are briefly discussed. However, various methods are not directly compared as different methods have different applications as well as concepts in their models.

Electric distance-based methods consider the fundamental characteristics of a power grid and the power flow while defining interactions among components. These graphs differ from data-driven interaction graphs as they directly use the inherent electrical properties of power grids based on Kirchoff's and Ohm's laws. They are also not specific to cascading failure studies and can be used for a variety of applications such as contingency analysis \cite{poudel2018electrical}, reliability studies for defining zones for load deliverability analysis \cite{blumsack2009defining}, response to targeted attacks \cite{bompard2012extended}, and so on.
Specifically, grid attributes such as admittance/impedance matrices, Jacobian matrices, and PTDF's reflect the operational features of power grids and can be used for the mentioned applications. Since these graphs do not rely on extensive simulation data, computational complexity of finding such graphs is relatively small.  

For cascading failure analysis, LODF-based interaction graphs \cite{guo2017monotonicity, guo2018failure} as well as interaction graphs showing the impact of hidden failures \cite{ma2017speeding, ma2017fast, ma2018fast} on lines during cascades have been used. These methods usually work well for cascade scenarios involving a small number of contingencies and smaller sized system. However, constructing such interaction graphs for cascade studies involving a large number of contingencies is computationally expensive. 
To this end, effective resistance-based interaction graphs can be used for cascade studies, as they can be easily produced by finding the effective resistance \cite{koc2014structural, kocc2014impact, wang2015network} between components without extensive power flow simulations. 
In addition to the above mentioned applications, electric distance-based interaction graphs can also be used for visualizing distinct electrical structure of power grids in two dimensions \cite{cuffe2015visualizing}.

\section{Reliability Analysis Using Interaction Graphs of Power Grids}
\label{Section4:ReliabilityAnalysis}
Interaction graphs constructed in Section \ref{Section3:InteractionGraphs} can be used for various analysis; specifically related to the reliability of the power grid, including analyzing the role of components and finding critical ones that contribute heavily in a cascade process, predicting distribution of cascades sizes, and studying patterns and structures that reveal connections and properties of the components in the power grid that extend beyond physical topology-based graphs. 
Thus, reliability studies performed using these interaction graphs are divided into various categories as discussed below.

\subsection{Critical Component Analysis}
The studies that identify and analyze the role of critical components in power grid's reliability are classified into three broad categories that include (1) using pre-existing as well as novel measures to find critical buses/transmission lines, (2) evaluating attack strategies that cause significant damage in the power grid, and (3) employing mitigation measures such as upgrading transmission lines or adding new components to protect the identified critical components.

\subsubsection{Critical Component Identification}
This class of reliability analyses focus on finding critical buses/transmission lines by analyzing structural properties of interaction graphs using standard centrality measures such as degree, betweenness, etc. (for a review of standard centrality measures, refer to \cite{borgatti2006graph}) or by defining novel interaction graph based metrics. 

\paragraph{Critical Component Identification Using Standard Centrality Measures}
In the studies presented in \cite{wei2018novel, wei2018complex, wei2019identification, zang2019adjacent, yang2019graph, wei2019electrical}, fault chain-based interaction graphs are found to be scale free graphs, indicating that most nodes possess low degrees but a limited number of nodes possess high in and out degrees. Thus, in the fault chain-based interaction graphs, vertices with higher degrees are assumed to be the critical components of the system.
Similar conclusions are obtained by authors in the studies in \cite{hines2008centrality, wang2011electrical}, where the impedance-based interaction graph is observed to be scale-free and consisting of a limited number of nodes with high degrees, which are considered as the critical components of the system. These are examples of works that consider the degree centrality measure to identify critical components of the system. Other centrality measures such as betweenness, eigenvector, and PageRank have also been considered on interaction graph-based representations of power grids including \cite{ma2017fast, ma2018fast, caro2015centrality} to find critical components of the system. 

\paragraph{Critical Component Identification Using New Centrality Measures}
\label{New Centrality Measures}
In addition to the studies that rely on standard centrality measures; some works develop new centrality measures in the context of power grids and the developed interaction graphs to analyze criticality of the components. For instance, in the generation-based interaction graphs \cite{qi2015interaction, ju2015simulation, ju2017multi, luo2017identify, qi2018efficient}, out-strength measure, which is the sum of the weights of the interaction links originating from a node, is used to find critical transmission lines. Such lines are the ones whose failure at any stage of the cascade including the initial stage or the propagation stage induces failure in a significant number of other transmission lines.
Outages in the initial stages are caused by external factors such as bad weather conditions, improper vegetation management, and exogenous events, whereas outages in the propagation stage is  due to power flow re-distributions, hidden failures, and other interactions between components as discussed in Section \ref{Section2:CascadingFailures}.
Influence-based \cite{hines2017cascading, zhou2018can} and multiple and simultaneous failure \cite{zhou2019markovian} interaction graphs are also used to find critical transmission lines, but they explicitly focus on lines whose failure during the propagation stage of cascading failures causes large cascades. Particularly, the studies in \cite{hines2017cascading, zhou2018can} use a cascade probability vector derived using the influence-based interaction graph to quantify the probability of failure of lines during the propagation stage of cascades and define critical lines as the ones whose corresponding entries in the probability vector have higher values. Similarly, the study in \cite{zhou2019markovian} finds the probability distribution of states of the multiple and simultaneous failure-based interaction graph and defines critical lines as the ones that belong to states with higher probability of occurrence.
Influence-based and correlation-based interaction graphs constructed in the studies in \cite{nakarmi2018analyzing, nakarmi2019critical} are also used to find the critical transmission lines during cascade processes by using a {\it community-centrality} measure. As the name suggests, the~measure quantifies the criticality of transmission lines based on their community membership, where critical lines are the ones that belong to multiple communities or act as bridges between communities. 
Note that communities are defined as groups of vertices with strong connections among themselves and few connections outside (for definition of communities and a review of community detection methods on graphs, refer to \cite{fortunato2016community}).

Identification of critical lines is not limited to data-driven interaction graphs. Multiple studies use electric distance-based interaction graphs for such analysis as well.
Effective resistance between components in the effective resistance-based interaction graph can be summed for all node pairs in the graph to find the {\it effective graph resistance} metric of the power grid. 
Effective graph resistance metric was initially defined in the study in \cite{klein1993resistance} as {\it Kirchhoff index} and used in the study in \cite{ellens2011effective} as a robustness metric.
Lower values of this metric suggest that the power grid is robust to cascading failures. 
Effective graph resistance can also be found using the eigenvalues and eigenvectors of the Laplacian matrix of the grid \cite{gutman1996quasi}.
In the study in \cite{koc2014structural}, critical transmission lines are found by measuring the changes in effective graph resistance before and after the removal of the line. 
In a similar manner, impedance-based interaction graphs constructed in the study in \cite{arianos2009power} and \cite{bompard2009analysis} are also used to find critical transmission lines by measuring the changes in net-ability metric before and after the removal of a line. Net-ability reflects the performance of a network by quantifying the ability of a generator to transfer power to a load within the power flow limits.

\subsubsection{Studying the Effect of Line Upgrades and Line Additions on Reliability of Power Grids}
While identification of critical components in the power grid is  necessary, assessing the impact of modifications and protection of such critical components in the overall power grid is the next step in the study. 
In the studies presented in \cite{hines2017cascading} and \cite{zhou2018can}, an influence interaction graph-based metric is used to quantify the impact of upgrading the critical lines (for example, by improving vegetation management around the lines or by improving protection systems) on cascade propagation.
The~work in \cite{zhou2019markovian} uses the multiple and simultaneous failure-based interaction graph to do a similar study.
In~both interaction graphs, the authors conclude that upgrading lines that take part in the propagation of failures during cascades reduces the risk of large cascades compared to the upgrade of lines that initiate cascades.
While the studies in \cite{hines2017cascading, zhou2018can} investigate the performance of the power grid networks after line upgrades, the studies in \cite{kocc2014impact} and \cite{wang2015network} use effective graph resistance metric to study the impact of adding transmission lines in optimal locations of the power grid.
However, in \cite{wang2015network} the authors warn that placing an additional line between a pair of nodes does not necessarily imply increased robustness of the grid. In fact, grid robustness may decrease after adding additional lines (due to Braess's paradox \cite{steinberg1983prevalence}), if the additions are done haphazardly.

\subsubsection{Analyzing Response to Attack/Failure Scenarios}
In addition to identifying critical components and characterizing the impact of their modifications in the reliability of power grids, the study of the response of power grids to attacks and failures is also necessary.
Such studies can be used to find critical components and attack strategies that threaten the reliability of the overall power grid.
In the studies in \cite{zhu2013risk, sun2014revealing, zhu2014resilience}, node integrated risk graphs are used to find groups of transmission lines whose removal from the graph causes the largest drop in net-ability of the power grid, as discussed in Section \ref{RiskGraphs}. These groups can be found in real-time independent of system parameters. The study in \cite{wang2015network} also analyzes the robustness of power grids to deliberate attacks using the effective graph resistance metric, as discussed in Section \ref{New Centrality Measures}.

\subsection{Prediction of Cascade Sizes}
Forecasting cascade sizes is another challenge in the reliability analysis of power grids. In the study in \cite{yang2017vulnerability}, a correlation graph-based statistical model, known as the {\it co-susceptibility} model, is used to predict cascade size distributions in transmission network of power grids using individual failure probabilities of transmission lines as well as failure correlations between transmission lines found from the correlation matrix. The study exploits the idea that groups of components that have higher correlations are likely to fail together and uses the correlation matrix to find such co-susceptible groups, which is an approximate estimate of the cascade size given an initial trigger failure. A similar idea is used in the studies in \cite{nakarmi2018analyzing, nakarmi2019critical}, where components within the same community are assumed to be likely to fail together, and the size of communities gives an approximation of cascade sizes.
The study in \cite{zhou2019markovian} also characterizes size of cascades by using the states of the Markov chain to find the probability distribution of the number of generations in a cascade.

\subsection{Studying Patterns and Structures in Interactions}
Structures and patterns in networks are important in describing the spread of various processes such as infectious diseases, behaviors, rumors, etc. \cite{anderson1992infectious, centola2010spread, daley1964epidemics}. For instance, in the studies \mbox{in \cite{nakarmi2018analyzing, nakarmi2019critical, guo2017monotonicity, guo2018failure}}, structures present in the interaction graphs of power grids are used to study the impact of cascading failures in the transmission network and utilize the graph structure to mitigate large cascades. The~graph structure considered in the studies in \cite{nakarmi2018analyzing, nakarmi2019critical} are communities whereas the graph structure considered in the studies in \cite{guo2017monotonicity, guo2018failure} are tree partitions.

In the study in \cite{guo2017monotonicity}, tree structures present in the outage condition-based interaction graph showed that transmission line failures could not propagate across common areas of tree partitions. Further, the extended work of \cite{guo2017monotonicity} in \cite{guo2018failure} found the critical components of the tree partitions, known as bridges. The failure of bridge lines plays a crucial role in the propagation of cascading failures. 
However, failure of non-bridge components does not propagate failures and the impact is more likely to be contained inside smaller regions/cells. This important property of bridge lines is used in mitigation of cascading failures by switching off transmission lines that cause negligible network congestion as well as improve the robustness of the system. 
Similarly, in the studies in \cite{nakarmi2018analyzing, nakarmi2019critical}, influence-based and correlation-based interaction graphs were used to limit propagation of cascading failures inside community structures. Particularly, the authors used the idea that failures can be trapped within communities by protecting the bridge/overlap nodes, which connect multiple communities together.

Purposes of analyzing structures present in the interaction graphs are not limited to mitigation of cascading failures. For instance, in the study in \cite{poudel2018electrical}, network structures were used for contingency analysis. The impedance-based interaction graph in \cite{poudel2018electrical} was pruned by removing edges above an operator defined threshold. Then, the common structure between the pruned impedance-based interaction graph and the topological graph of the power grid was analyzed and verified to be the contingencies that violate transmission line limits and cause overloads. 
Similarly, the study in \cite{blumsack2009defining} identified zonal patterns in the impedance-based interaction graph for reliability assessment of zones for load deliverability analysis.

\section{Summary}
\label{Summary}
Analyzing interactions among the components of power grids during cascading failures can reveal important information about the failure propagation process in these systems and the role of the components in the process. It was discussed that the physical topology of the power grids have limitations in modeling cascading failure process in power grids. As such, new methods are emerging to capture non-local influences/interactions among the components of the power grid. In this survey, various techniques for constructing the graph of interactions (beyond the physical topology of the power grid) were reviewed. A novel taxonomy was presented for classifying the existing research studies into various categories and subcategories based on the type of data and techniques used in inferring the interactions and creating the graph model of the system. Finally, a brief
overview of various reliability analysis studies based on the interaction graphs was presented.

Specifically, the presented taxonomy introduced two main categories for the methods of constructing interaction graphs: data-driven and electric distance-based approaches. The key properties and limitations of various techniques in each category as well as an overview of the methods in each category have been discussed. While for data-driven methods, availability of cascade data is a necessity, the electric distance-based methods rely on accurate physical models of the system (which are not required for data-driven methods). Therefore, depending on the availability of the data or the physical model of the system, one can choose between these categories of methods. Moreover, as the existing methods use various information for inferring and modeling interactions, they can shed light on various aspects of the cascade process and reliability and vulnerability of the components. As such, depending on the analysis of interest, different techniques can be selected.
Studies in both areas of data-driven and electric distance-based approaches are ongoing, and it can be expected that these methods will greatly contribute in providing new understandings about the cascade process in power grids.
Furthermore, in the area of cascading failure reliability analysis based on interaction graphs, recent literature is more inclined towards identifying structures and patterns in the interaction graphs to study the vulnerabilities in power grids. However, other analyses, such as predicting the path and size of the cascades, can also utilize the interaction graphs of the system.

 {Next, future directions of research related to the interaction graphs for cascading failure analysis are briefly reviewed.}
One of the key areas of research in this domain would be to relax the dependence of the constructed interaction graph on the applied method, data, and operating characteristics of the system. As discussed in this survey, the method used for inferring the interactions; the available data; and the operating setting of the system, such as the loading level of its components, all affect the interaction graphs. It is important to be able to derive unified graphs of interactions that implicitly capture the effect of such factors in the model. 
Moreover, the computational complexity of constructing  interaction graphs (which increases with the size and scale of power grids) has limited their application to off-line analysis of reliability. This limitation suggests the need for further research in developing more effective data-driven and electric distance-based methods for developing real-time analysis based on interaction graphs for time sensitive analysis during cascading failures.

In addition, most of the existing studies using the interaction graphs are observed to be focused on the identification of the critical components of the power grid. Such analyses can be a step towards developing mitigation strategies that can be applied to create more reliable power grids. As such, it~is important that practical and economically feasible mitigation strategies be studied and developed, based on the insights provided by the interaction graphs.

Finally, in order to create resilient and reliable power grids, it is important to understand the role of interactions among the components of the system in reliability challenges, such as cascading failures, as discussed in this paper. However, in addition to the reliability analysis of the physical components of the system, it is also important to consider soft factors, such as human factors and their influences on the system as well as operating policies in reliability analysis and planning. Such~interaction analyses will be an essential research direction for future studies of cascading failures and one example of such study is presented in the work in \cite{wang2018impacts}.


\ifCLASSOPTIONcaptionsoff
  \newpage
\fi

\bibliographystyle{IEEEtran}
\bibliography{ReferencesFile}

\end{document}